\documentclass{article}

\usepackage{arxiv}

\usepackage[utf8]{inputenc} 
\usepackage[T1]{fontenc}    
\usepackage{hyperref}       
\usepackage{url}            
\usepackage{booktabs}       
\usepackage{amsfonts}       
\usepackage{nicefrac}       
\usepackage{microtype}      
\usepackage{lipsum}		
\usepackage{graphicx}
\usepackage{float}

\title{Twitter Sentiment Analysis using Distributed Word and Sentence Representation}

\author{
  Dwarampudi Mahidhar Reddy\\
  Computer Science and Engineering Department\\
  Manipal Institute of Technology\\
  Manipal, Karnataka, 576104, India \\
  \texttt{mahidhar\_d@hotmail.com} \\
  \And
  Dr. N V Subba Reddy \\
  Computer Science and Engineering Department\\
  Manipal Institute of Technology\\
  Manipal, Karnataka, 576104, India \\
  \texttt{nvs.reddy@manipal.edu} \\
  \And
  Dr. Prema K V \\
  Computer Science and Engineering Department\\
  Manipal Institute of Technology\\
  Manipal, Karnataka, 576104, India \\
  \texttt{prema.kv@manipal.edu} \\
}

\begin{document}
\maketitle

\begin{abstract}
An important part of the information gathering and data analysis is to find out what people think about, either a product or an entity. Twitter is an opinion rich social networking site. The posts or tweets from this data can be used for mining people’s opinions. The recent surge of activity in this area can be attributed to the computational treatment of data, which made opinion extraction and sentiment analysis easier. This paper classifies tweets into positive and negative sentiments, but instead of using traditional methods or preprocessing text data here we use the distributed representations of words and sentences to classify the tweets. We use Long Short-Term Memory (LSTM) Networks, Convolutional Neural Networks (CNNs) and Artificial Neural Networks. The first two are used on Distributed Representation of words while the latter is used on the distributed representation of sentences. This paper achieves accuracies as high as 81\%. It also suggests the best and optimal ways for creating distributed representations of words for sentiment analysis, out of the available methods.
\end{abstract}


\section{Introduction}
Twitter is a microblogging site which has involved to become a source of opinion rich information. It is because people post real time messages or tweets about their opinions on a variety of entities and issues. This data can be used to find the opinion or sentiment of people towards an entity, issue or a product. This information can be later used to either address the issue or make improvements to a product or make changes to an entity. In many cases these tweets are monitored and the users are addressed almost immediately. The challenge here is to summarize the sentiment of the whole tweet and automate this task.
\par
In this paper we try to classify tweets into positive and negative sentiments. We experiment with three types of Neural Networks, Multi-layer perceptron, LSTMs and CNNs, the former is used for distributed representation of Sentences or sentence vectors and the other two are used on the distributed representation of words. We also experiment with three methods of building word vectors Word2Vec Continuous Bag of Words (CBOW), Word2Vec Skipgram and FastText. to find the best one and optimal methods for sentiment analysis.
\par
We built a Naïve Bayes model as a base model with traditional preprocessing methods to compare it with the models built trained on word and sentence vectors. These models are completely trained on text, none of the non-word tokens or symbols were considered. A complete section is dedicated to preprocessing later in the paper where everything is discussed in detail.

\section{Literature Survey}
Sentiment analysis is usually handled as Natural Language processing(NLP), In ‘Sentiment Analysis of Twitter Data’ by Apporv Agarwal, Boyi Xie , Ilia Vovsha, Owen Rambow, Rebecca Passonneauto\cite{5}, they introduced POS-Specific prior polarity features. In ‘Comparison of Text Classification Algorithms’ by M. Trivedi, S. Sharma. N. Soni, S. Nair\cite{6}, they compared various algorithms such as support vector machines (SVM) and C4.5 and found out that SVM was the better performing model. In ‘A review on Feature Selection and Feature Extraction for Text Classification’ by Foram P. Shah and Vibha Patel\cite{7}, they discuss a lot of methods to extract features from text documents for classification. We talked about all these text classification task because, sentiment analysis is a in a way a subset of text classification. In text classification we classify text into predefines categories, while in Sentiment Analysis those categories are Sentiments.
\par
\begin{figure}[H]
	\centering
	\fbox{ \includegraphics[scale=0.25]{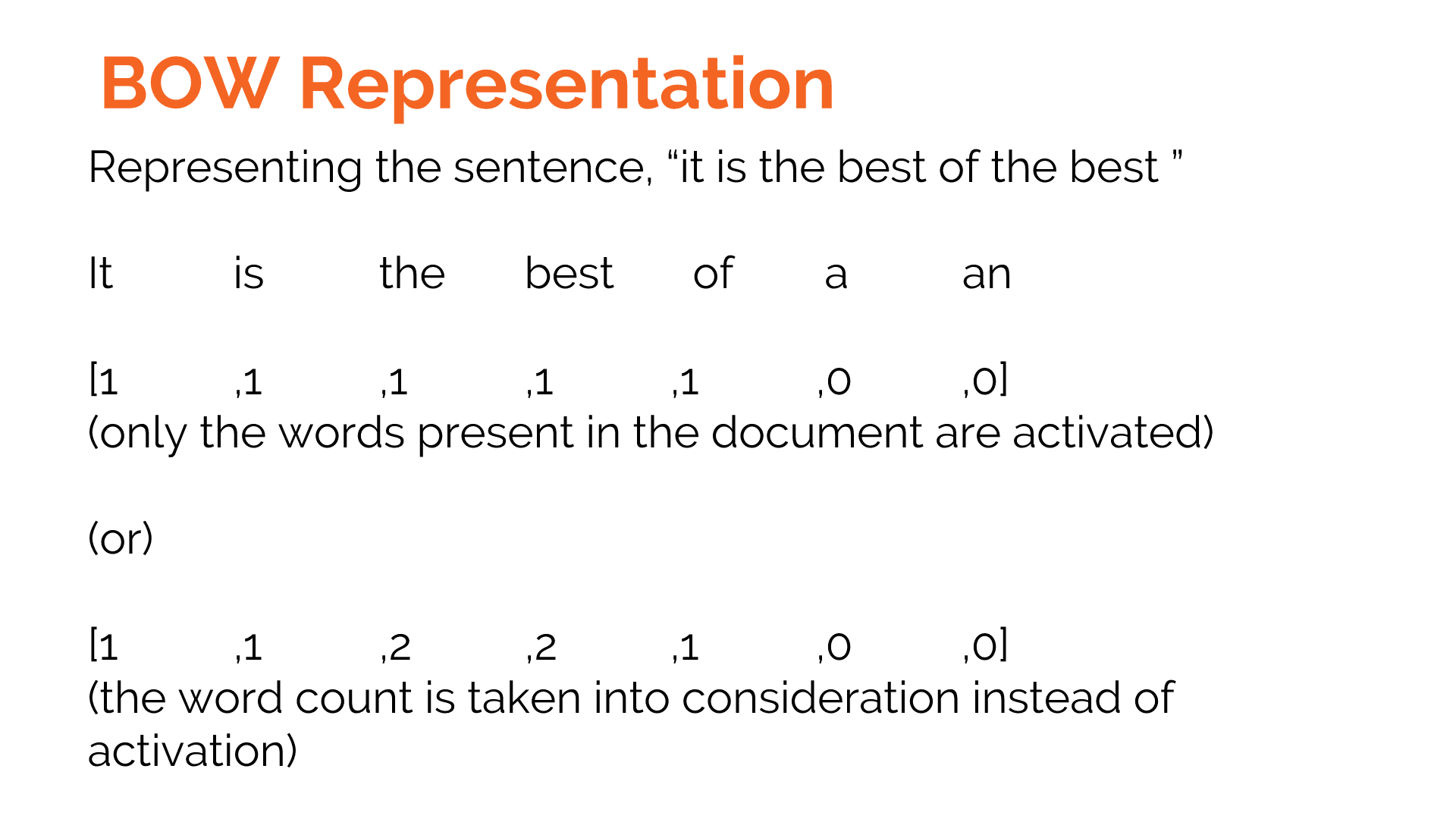}}
	\caption{Bag of Words representation of the sentence “It is the best of the best”, where every word in the sentence is considered as a feature, if only a few words are chosen as feature words, then the rest of the words are not even represented in the vector.}
	\label{fig:bow_representation}
\end{figure}
\begin{figure}[H]
	\centering
	\fbox{ \includegraphics[scale=0.25]{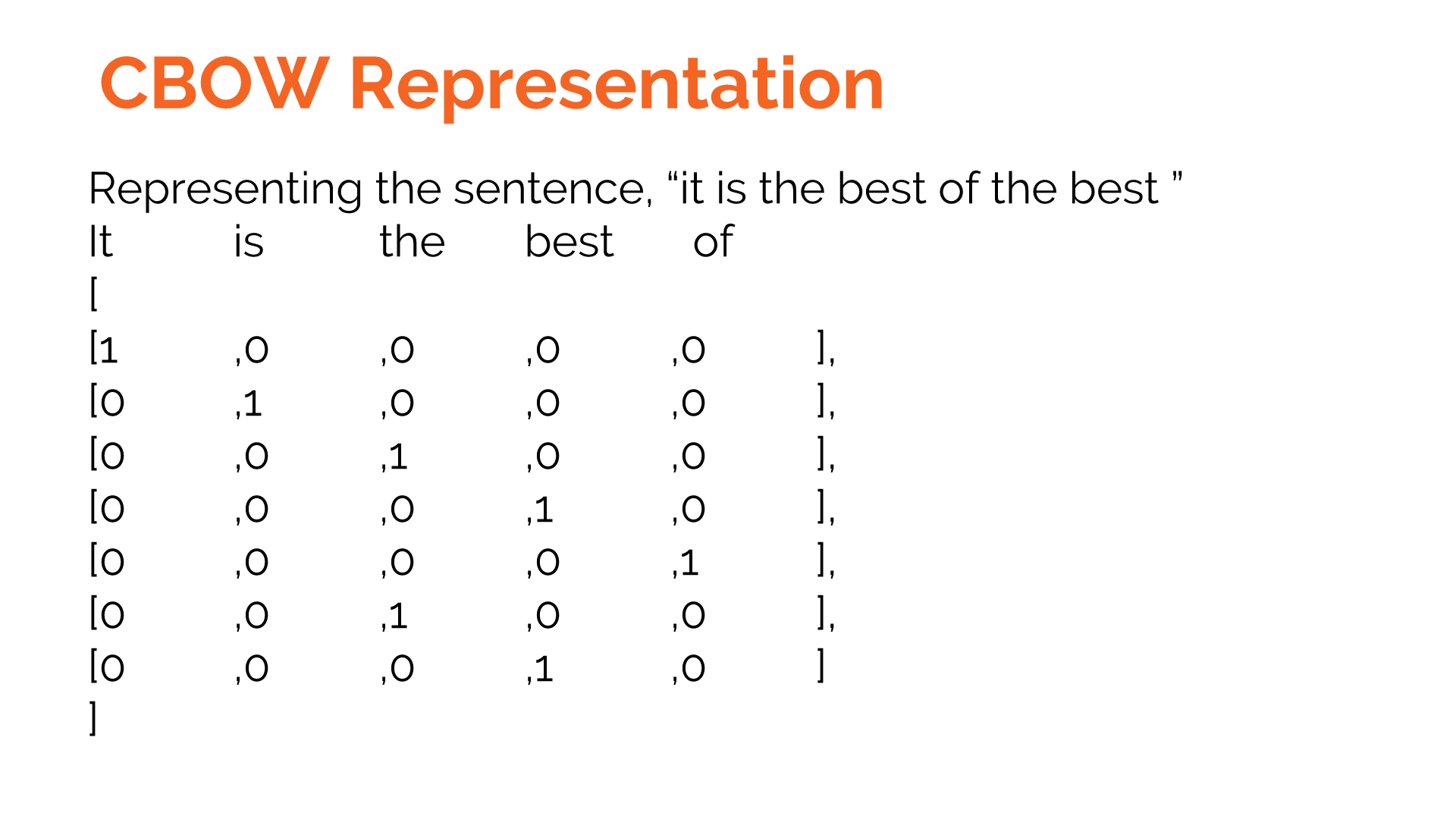}}
	\caption{Continuous Bag of Words representation of the sentence “It is the best of the best”. As you can see every word has its own vector inside the sentence vector, which has only one activated element and the rest are zeros, as the number of words increase the size of each of these naïve word vectors increase.}
	\label{fig:cbow_representation}
\end{figure}

In all the above-mentioned papers, words were used as features from the documents. As we know that many machine learning algorithms require the data to be represented as a fixed length feature vector. There are 2 very common ways of representing sentences when words are used as features, Bag of Words(BOW) representation and Continuous bag of words (CBOW). Figure \ref{fig:bow_representation} shows the representation of the sentence “It is the best of the best. “. The problem with Bag of Words Representation is it loses the order and meaning of the words. To over some this the CBOW model was proposed where the words are activated according to their occurrence in the sentence. Figure \ref{fig:cbow_representation} shows CBOW representation of the same sentence. CBOW has problems with memory, as the sentence size increases the length of the sentence vector also increases, to cover most the words, each word should be a feature, this increases the size of the sentence vector drastically. Moreover, these vectors contain only once activated row and the rest are zeros.

Apart from the memory problem, both CBOW and BOW has problem with missing words, as not all words can be represented in the vector, only a few are taken using various methods \cite{7}, the rest of the words which might have significance are lost. Hence, we need a newer representation.

So instead of using words as features, we can use word vectors which have fixed length and won’t change irrespective of the number of words in the dataset. This way we can efficiently use the memory. There are 3 different ways of constructing word vectors, Word2Vec CBOW \cite{1}, Word2Vec skipgram \cite{2} and FastText \cite{4}. Word2Vec CBOW uses a window of words around a word, there will be n input layers for the neural network with all the words are inputs, only the words around it will be activated in the layer according to the distance of the input from the actual word and the actual word will be the output. While training the weights are back propagated, then the whole neural network is reversed where the output layer becomes the input layer and now in the reversed network the output layer is removed and the output from the penultimate layer is taken as the word vector. Skipgram is slightly different, the input word is the word for which we want to find the vector and the output word will be the words around it in the window, after training the last layer is removed as usual and the penultimate layer’s output is considered as the word vector. Document vectors or Sentence Vectors are also made in the same way but a sentence id is added additionally to the sentence. FastText is slightly different, instead of considering the word as atomic, it considers the n grams and trains on them and, when added they give the word vector.
\par
Moreover, Distributed word representations can identify and relate to typographical errors and internet slang.
\par
Here we use these word vectors and a modified CBOW representation, where instead of activating the word, we replace it with the word vector for the sentence.

\section{Data}
The dataset used in this paper were published on the Thinknook website \cite{3}. Where the maximum number of words in a tweet is 93. There where are total of 1,578,612 tweets. Only 10\% of the tweets were used for all the experiments in this paper, i.e., 157,860 tweets. The train test split was 7:3. The In the train data the number of positive sentiment tweets are, 63,001 and the number of negative tweets is 63,287, making the total number of tweets in the train set to be 126,288, the rest, i.e., 31,572 in the test set. A few sample tweets are displayed in Table ~\ref{tab:unprocess_sam}.

\begin{table}[H]
	\caption{Sample tweets from the dataset}
	\centering
	\begin{tabular}{lll}
		\toprule
		
		& Tweet     & Sentiment \\
		\midrule
		& @scoutout You reminded me of Ralph S. Mouse. If you read Beverly Cleary, you know who I'm talking about.  & 1   \\
		& cryinnnnng---------- so sadd ........... .. i cant lifee.............i hate my lifeee...!! i can't stop....!!! plzz i need helpp & 0     \\
		& my cat is happy so she's drooling from 3 places of her mouth. I love felix       & 1  \\
		& @joyce\_ap ive been to cebu and bohol na.       & 0  \\
		& @RHEAAAxx i feel cool. i'm the only non asian person you're following.       & 1  \\
		& i have nothing to do until thursday WHYY OH WHYY MEEEE?!       & 0  \\
		&http://www.bizjournals.com/portland/stories/2009/06/01/daily38.html maybe it's time to start looking east       & 0  \\
		& http://twitpic.com/65v9u - Awwww well aren't we just hella cute       & 1  \\
		\bottomrule
	\end{tabular}
	\label{tab:unprocess_sam}
\end{table}

\subsection{Preprocessing}
The Several Steps were carried out in preprocessing the tweets. This does not include creating testable data on which the model is trained. The first part of preprocessing the data is cleaning the data. It includes 3 steps:

\begin{itemize}
	\item Replacing Numbers: All the numbers occurring in the tweets are replaced by ‘0’ in the string.
	\item Replacing all the twitter handles: All the twitter handles in the tweets are replaced with ‘1’. Twitter handles are usually in the form of “@handle”, with alpha numeric characters. 
	\item Replacing all the URLs: All the URLs in the tweets are replaced with ‘2’.
\end{itemize}

This was the first part of the preprocessing, The Tweets in table 1 are displayed after preprocessing in Table ~\ref{tab:process_sam}.

\begin{table}[H]
	\caption{Tweets after first part of preprocessing}
	\centering
	\begin{tabular}{lll}
		\toprule
		
		& Tweet     & Sentiment \\
		\midrule
		& 1 you reminded me of ralph s. mouse. if you read beverly cleary, you know who i'm talking about.  & 1   \\
		& cryinnnnng---------- so sadd ........... .. i cant lifee.............i hate my lifeee...!! i can't stop....!!! plzz i need helpp & 0     \\
		& my cat is happy so she's drooling from 0 places of her mouth. i love felix       & 1  \\
		& 1 ive been to cebu and bohol na.       & 0  \\
		& 1 i feel cool. i'm the only non asian person you're following.       & 1  \\
		& i have nothing to do until thursday whyy oh whyy meeee?!      & 0  \\
		&2 maybe it’s time to start looking east      & 0  \\
		& 2 – awwww well aren’t we just hella cute       & 1  \\
		\bottomrule
	\end{tabular}
	\label{tab:process_sam}
\end{table}

After this we tokenize the sentences, every word becomes a token. All the non-word tokens are omitted and only the text is considered for this task. After tokenization, all the words are lemmatized \cite{11} and the repetitions are removed for unique words. Stop words are not removed as negative stop words such as “hasn’t” effect the sentiment of the tweet \cite{8}.
\par
These are used for training the word vectors and sentence vectors, which are later used to train the models.

\subsection{Construction of distributed representation of words and sentences}
\begin{figure}[H]
	\centering
	\fbox{ \includegraphics[scale=0.8]{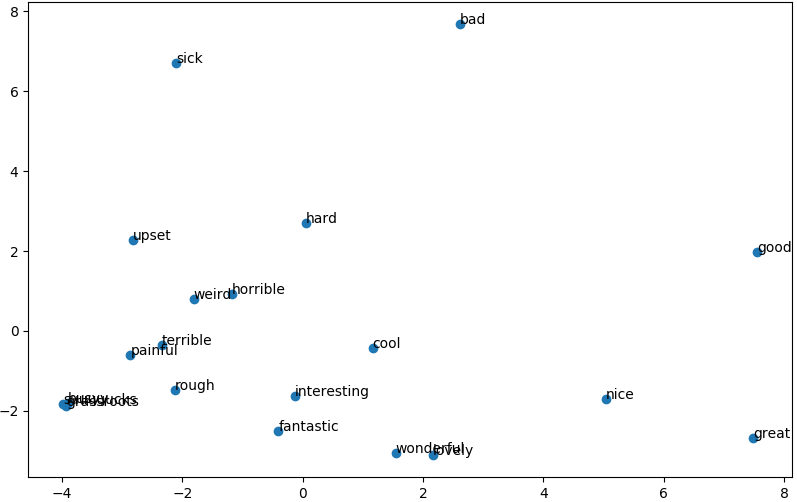}}
	\caption{Word2Vec CBOW representation of words are plotted in this graph.}
	\label{fig:w2v_cbow}
\end{figure}

Here we construct distributed representation of words and sentences and express them in their final form which is used to train the models. We use a third-party library, gensim \cite{9}, for this purpose, we constructed vectors for all the words and the 3 types of construction. The word vectors for a total of 251,292 tokens were generated. Here are the 3 plots for plotting the most similar words for the words ‘good’ and ‘bad’. PCA is used for dimensionality reduction. Figure \ref{fig:w2v_cbow}, shows the plot for Word2Vec CBOW. Here you can see that all the negative words or the words related to ‘bad’ are slightly on the top and the words related to ‘good’ or positive words are slightly at the bottom, but the converge as we move to the left.

\begin{figure}[H]
	\centering
	\fbox{ \includegraphics[scale=0.8]{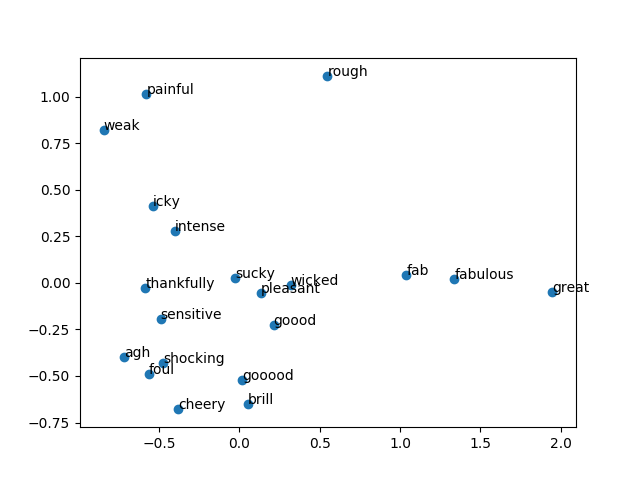}}
	\caption{Word2Vec Skipgram word vectors are plotted in this graph}
	\label{fig:w2v_sg}
\end{figure}

Figure \ref{fig:w2v_sg}, shows the skipgram graph, where you can see the the words related to ‘good’ are better distinguished from the words related to ‘bad’, but even here you see them mixing, as in ‘gooood’, ‘cheery’ are very close to ‘shocking’ and ‘foul’. Similarly, ‘pleasant’ and ‘goood’ are very close to ‘sucky’ and ‘wicked’.

\begin{figure}[H]
	\centering
	\fbox{ \includegraphics[scale=0.5]{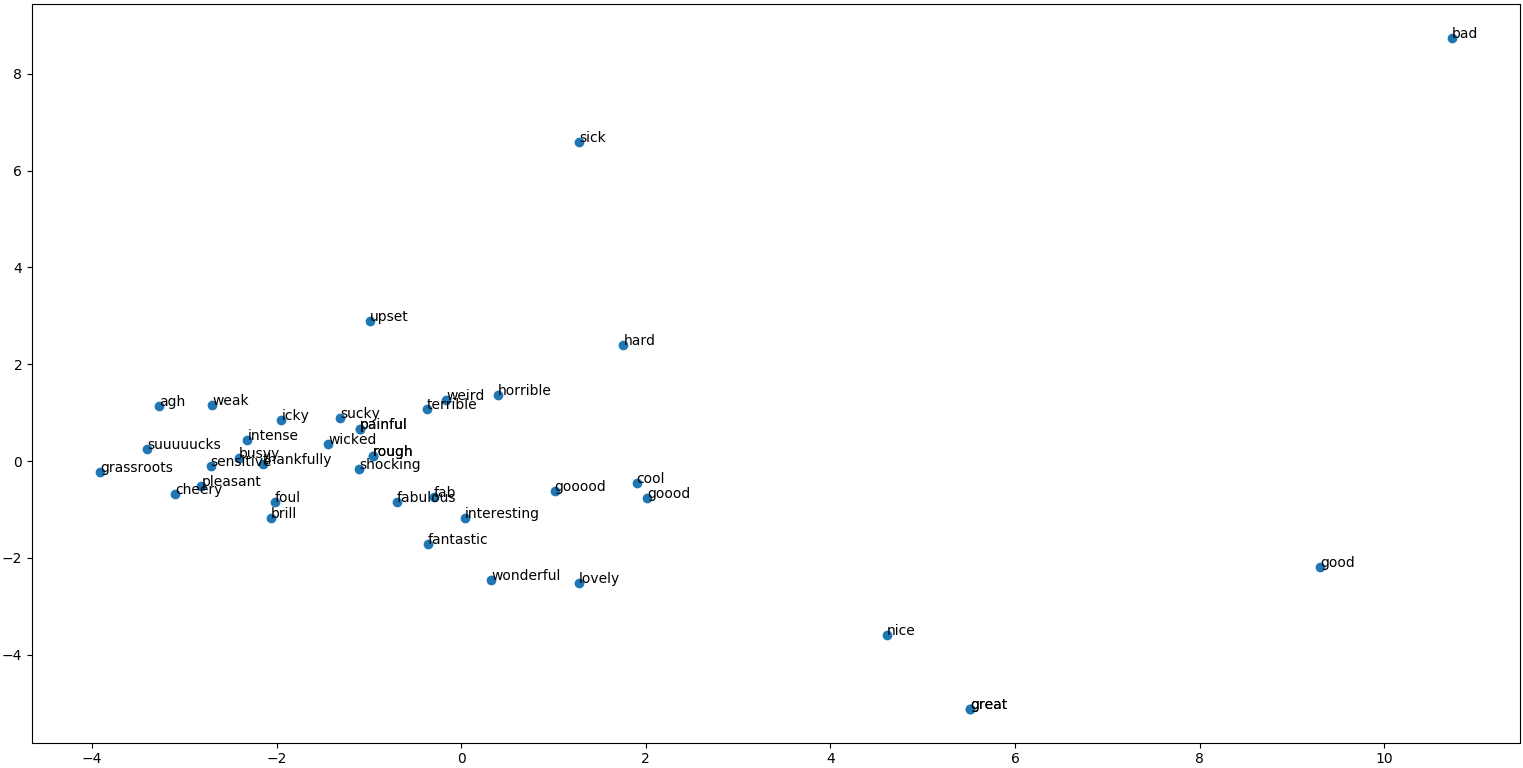}}
	\caption{FastText word vectors are plotted in this graph. Here, an almost clear distinction between the words related to ‘good’ and ‘bad’ can be seen though they become very close at times. Such words are used in both positive and negative contexts.}
	\label{fig:w2v_ft}
\end{figure}

Figure \ref{fig:w2v_ft}, shows the FastText graph, since fasttext doesn’t store any word vectors, the words from CBOW and Skipgram are used from plotting on FastText graph, where the vectors were constructed using the n gram vectors of FastText.
\par
From, these figures we can infer that, FastText should be the most accurate model and CBOW is the least while representing words. But we further compare them on 2 different artificial neural networks, Long Short-Term Memory Networks and Convolutional Neural Networks.
\par
For the Sentence Vectors we use the Normal, Multi-Layer Perceptron. The Sentence Vectors were trained on multiple dimensional lengths, that is, 100, 200, 300, 400 and 500, to check if the number of dimensions effect the accuracy of the model.
\par
Similarly, Word Vectors for the optimal Vectors were also trained for multiple lengths to check for the change in accuracies for the change in number of dimensions. The input vectors for the modified CBOW are padded with using the pre-padding techniques where the padding is done before the data, as the models take fixed length input.

\section{Classification}
The Classification is mostly done on LSTMs \cite{10} \cite{14} and CNNs \cite{13} \cite{12} \cite{15} but just for the sake of comparison, we compare it with Naïve Bayes model. For the Naïve Bayes Model, Laplace smoothing or Add-One smoothing is applied so that, whenever an unseen word occurs it doesn’t throw a zero probability.
\par
Two different LSTMs \cite{10} were used in this paper one with only one hidden layer (LSTM-1) and the other with 4 hidden layers (LSTM -4), the number of neurons on all the hidden layers are 100. The number of input neurons on the network change with word dimensions. Tanh was used as the activation function for al the layer except the last classification layer where sigmoid was used. Dropout of 0.2 was used on every layer.
\par
The Convolutional Neural Networks [15] Used in this paper are also in parallel with the LSTMs, 2 were used, one hidden layer (CNN – 1) and 4 hidden layers (CNN - 4). With varied input layer neurons according to the word dimensions. Linear activation function on every layer and dropout of 0.2 was applied. The last layer uses a sigmoid activation.
\par
Multilayer perceptron of 3 layers were used Sentence Vectors. With a dropout of 0.2 on all the layers and rectified linear units or ReLu was used as activation on all the layers, except the last one where sigmoid was used.

\section{Results}

The following tables compare the results, accuracies between all the above-mentioned model and types of word vector construction, and as promised tells us the most optimal was for constructing the vectors and the most accurate way for constructing the vectors.

\begin{table}[H]
	\caption{LSTM-1 vs LSTM-4 vs CNN-1 vs CNN-4, all with 100 Neurons on each Layer (\%)}
	\centering
\begin{tabular}{lllll}
	\toprule
		& LSTM-1 	& LSTM-4 	& CNN-1 	& CNN-4 \\
	\midrule
	Train & 81.663  & 80.072 	& 74.746  	& 74.721    \\
	Test  & 80.175 	& 80.321   	& 74.756 	& 74.993     \\
	Epochs & 9    	& 9 		& 7  		& 10 \\
	\bottomrule
\end{tabular}
	\label{tab:lstm_cnn}
\end{table}

Table ~\ref{tab:lstm_cnn} has the results or skipgram model on various models. Since the results of 4 layered networks are higher than the single layered networks, the rest of the datasets were tested on the 4 layered networks. Also, as we can see that LSTM performed significantly better than the CNN. The rest of the datasets were only tested on LSTM with 4 hidden layers.

\begin{table}[H]
	\caption{Skipgram vs CBOW vs FastText on LSTM-4 (\%)}
	\centering
	\begin{tabular}{llll}
		\toprule
		& Word2Vec Skipgram 	& Word2Vec CBOW 	& FastText 	\\
		\midrule
		Train & 80.072  & 79.641 	& 81.858  	    \\
		Test  & 80.321 	& 50.019   	& 80.622 	    \\
		Epochs & 9    	& 5 		& 15  		 \\
		\bottomrule
	\end{tabular}
	\label{tab:sgftcbow}
\end{table}

From Table ~\ref{tab:sgftcbow}, Though FastText is the best performing model, it’s performance is only slightly higher than Skipgram. Hence, when time is also taken in to consideration, Skipgram is the best model. Hence Skipgram is taken for further testing. Also, CBOW quickly overfits after 5 epochs, even with a dropout of 0.3 on every layer.

\begin{table}[H]
	\caption{LSTM-4, word dimensions 100 vs 200 vs 300 (\%)}
	\centering
	\begin{tabular}{llll}
		\toprule
		& 100 dimensions 	& 200 dimensions 	& 300 dimensions 	\\
		\midrule
		Train & 80.072  & 82.682 	& 82.120  	    \\
		Test  & 80.321 	& 81.027   	& 80.892 	    \\
		Epochs & 9    	& 16 		& 12  		 \\
		\bottomrule
	\end{tabular}
	\label{tab:dims}
\end{table}

From Table ~\ref{tab:dims}, 200 dimensions have slightly higher accuracy but it decreased in 300 dimensions. Further investigation is required here and need to be tested with higher dimensional vectors. But 52GB memory wasn’t enough for computing the results.

\begin{table}[H]
	\caption{FeedForward Neural Network accuracy, Sentences Vector Dimensions of 100, 200, 300, 400, 500 (\%)}
	\centering
	\begin{tabular}{llllll}
		\toprule
		& 100 	& 200 	& 300 	& 400 & 500 \\
		\midrule
		Train & 67.949  & 67.961 	& 67.512  	& 68.186 & 66.942 \\
		Test  & 67.993 	& 68.399   	& 67.968 	& 68.431 & 67.544 \\
		Epochs & 38    	& 15 		& 6  		& 26 & 3 \\
		\bottomrule
	\end{tabular}
	\label{tab:ff}
\end{table}

From Table ~\ref{tab:ff}, Nothing conclusive can be drawn from results in Table ~\ref{tab:ff} as there is no constant increase in accuracy or decrease in the epochs but changing the shape of the network might show some results.

\section{Conclusion}
This paper provides an implementation of sentiment analysis on tweets using distributed representation of words and sentences and shows that, it is memory efficient theoretically. We concluded that LSTMs are better at Sentiment Analysis. Then we found out that higher the number of hidden layers will almost always give you greater efficiencies. Then concluded that word vectors are better for sentiment analysis than sentence vectors. We also concluded that Word2VEc skipgram is the most optimal way of generating distributed representation of sentiment analysis and FastText gives the highest accuracies but takes more time.

\bibliographystyle{unsrt}

\end{document}